\documentstyle[sprocl]{article}
\def\lsim{\mathrel{\rlap{\lower4pt\hbox{\hskip1pt$\sim$}}
    \raise1pt\hbox{$<$}}}         %less than or approx. symbol
\def\gsim{\mathrel{\rlap{\lower4pt\hbox{\hskip1pt$\sim$}}
    \raise1pt\hbox{$>$}}}         %greater than or approx. symbol

\input psfig.sty

\def\Journal#1#2#3#4{{#1} {\bf #2}, #3 (#4)}

\def\PR{\em Phys. Repts.}

\def\PRL{\em Phys. Rev. Lett.}
\def\PRD{{\em Phys. Rev.} D}

\def\ApJ{\em Ap. J.}
\def\ApJS{\em Ap. J. Supp.}
\def\Nature{\em Nature}
\def\MNRAS{\em M.N.R.A.S.}
\def\SA{\em Sov. Astr.}
\def\SAL{\em Sov. Astr. Lett.}
\def\ASS{\em Ap. Space Sci.}
\def\PRC{{\em Phys. Rev.} C}

\def\be{\begin{equation}}
\def\ee{\end{equation}}
\def\bea{\begin{eqnarray}}
\def\eea{\end{eqnarray}}
\begin{document}

\title{NEUTRINO EFFECTS IN NUCLEOSYNTHESIS}

\author{W. C. HAXTON}

\address{Institute for Nuclear Theory, Box 351550, and 
Department of Physics, Box 351560, \\
University of Washington, Seattle, WA 98195, USA \\
E-mail: haxton@phys.washington.edu}

%%%%%%%%%%%%%%%%%%%%%%%%%%%%%%%%%%%%%%%%%%%%%%%%%%%%%%%%%%%%%%
% You may repeat \author \address as often as necessary      %
%%%%%%%%%%%%%%%%%%%%%%%%%%%%%%%%%%%%%%%%%%%%%%%%%%%%%%%%%%%%%%

\maketitle\abstracts{ The nucleosynthesis within a Type II
supernova occurs in an intense neutrino flux.  I discuss some
of the effects associated with neutrino interactions, including 
direct synthesis in the neutrino process, the role of
neutrinos in controlling the r-process path and in postprocessing
r-process products, and neutrino oscillation connections.}

It is a great pleasure to attend this meeting in honor of a
long-time friend, Frank Avignone, and dedicated to his favorite
subject, neutrino physics.  In contrast to the rare neutrino events
that Frank has measured in the laboratory, I will talk about an 
environment where neutrino reactions are so frequent that they 
determine much of the chemistry of the matter.  That environment
is the progenitor star's envelope in
the first seconds after a core-collapse supernova.  Here the 
neutrinos directly synthesize new nuclei, and help to eject that
matter into the interstellar medium, where it is incorporated 
into stars like our sun.  The neutrinos also control the isospin
of the nucleon soup that is the likely site of the r-process.
It follows that there is an intimate connection between the 
properties of neutrinos, including phenomena like neutrino
oscillations, and supernova nucleosynthesis.

\section{Core-Collapse Supernovae}
In the infall stage of a core-collapse supernova~\cite{janka} neutrinos are
trapped by their neutral current interactions once a density
of $\rho \sim 10^{12}$ g/cm$^3$ is reached.  Trapped in this
sense means that the neutrino diffusion time becomes longer than
the time needed to complete the collapse, thereby guaranteeing
that the energy liberated by the matter falling into the
gravitational potential, $\sim 3 \times 10^{53}$ ergs, is 
contained within the protoneutron star.  A small portion of this
energy is later apparent in the kinetic energy of the ejected
shells and in the accompanying optical display.  But the vast
majority, $\sim$ 99\%, is radiated in neutrinos over the 
$\sim$ 3 second cooling time of the core, following core bounce.

Throughout most of their outward diffusion, the various neutrino
flavors remain in equilibrium
\begin{equation}
\nu_e +\bar{\nu}_e \leftrightarrow \nu_\mu + \bar{\nu}_\mu
\end{equation}
thereby ensuring that the energy is shared equally by the three
flavors.  However, when they reach the ``neutrinosphere'' at 
$\sim 10^{12}$ g/cm$^3$, their decoupling is flavor dependent
due to the reactions
\begin{eqnarray}
\nu_x + e &\leftrightarrow& \nu_x + e \nonumber \\
\nu_e + n &\leftrightarrow& p + e^- \nonumber \\
\bar{\nu}_e + p &\leftrightarrow& n + e^+ . 
\end{eqnarray}
The first reaction for $\nu_e$s is about six times that for 
heavy flavors, while the second and third affect only electron
neutrinos.  As a result the heavy-flavor neutrinos decouple at
a higher density, and thus temperature, than the electron
neutrinos.  The result is a characteristic temperature
hierarchy
\begin{eqnarray}
T_{\nu_\mu,\nu_\tau} &\sim& 8 \mathrm{MeV} \nonumber \\
T_{\bar{\nu}_e} &\sim& 4.5 \mathrm{MeV} \nonumber \\
T_{\nu_e} &\sim& 3.5 \mathrm{MeV}
\end{eqnarray}
where the $\nu_e - \bar{\nu}_e$ temperature difference results 
from the matter near the neutrinosphere being neutron rich (having
experienced significant electron capture).  As the energy is divided
approximately equally among the flavors, it follows that the 
electron neutrino flux is about twice that of the heavy flavors.

Supernovae are important engines driving galactic evolution, producing
and ejecting the metals that enrich the galaxy.  
Elements produced in the hydrostatic evolution of the
presupernova star (C, O, Ne, ...) are abundant in the ejecta
of the explosion.  The shock wave resulting from core bounce produces
peak temperatures of $\sim (1-3) \cdot 10^9$K as
it traverses the silicon, oxygen, and neon shells.  This
shock wave heating induces proton and $\alpha$ reactions like $(\gamma,\alpha) \leftrightarrow 
(\alpha,\gamma)$ which generate a 
mass flow toward highly bound nuclei, resulting in the
synthesis of iron peak elements as well as less abundant
odd-A species.  Rapid neutron-capture reactions are thought
to take place in the high-entropy atmosphere just above
the mass cut, producing about half of the heavy elements 
above A $\sim$ 80.  Finally, the neutrinos themselves transmute
certain nuclei within the mantel, producing rare isotopes like
$^{11}$B and $^{19}$F in the neutrino process.
  
\section{The Neutrino Process}
The neutrino process was described independently by Domagatsky 
{\it et al.}~\cite{domo} and by Woosley, Haxton, {\it et al.}~\cite{wh}  Probably the
simplest example occurs in the neon shell
in a supernova.  Because of the first-forbidden contributions,
the cross section for inelastic neutrino scattering to the 
giant resonances in Ne is $\sim 3 \cdot 10^{-41}$ cm$^2$/flavor
for the more energetic heavy-flavor neutrinos.
This reaction
\begin{equation}
\nu + A \rightarrow \nu' + A^* 
\end{equation}
transfers an energy typical of giant resonances, $\sim$ 20 MeV.
A supernova energy release of 3 $\times 10^{53}$ ergs
converts to about $4 \times 10^{57}$ heavy
flavor neutrinos.  The Ne shell in a 20 M$_\odot$ star has
at a radius $\sim$ 20,000 km.  Thus the neutrino fluence through
the Ne shell is
\begin{equation}
 \phi \sim { 4 \cdot 10^{57} \over 4 \pi (20,000 \mathrm{km})^2 }
\sim 10^{38}/\mathrm{cm}^2. 
\end{equation}
Thus folding the fluence and cross section,
one concludes that approximately 1/300th of the Ne nuclei interact,
often breaking up to form $^{19}$F.

This is quite interesting since the astrophysical origin of $^{19}$F
had not been understood.  The only stable isotope of fluorine,
$^{19}$F has an abundance
\begin{equation}
{^{19}\mathrm{F} \over ^{20}\mathrm{Ne}} \sim {1 \over 3100}. 
\end{equation}
This leads to the conclusion that the fluorine 
found in toothpaste was
created by neutral current neutrino reactions deep inside some
ancient supernova. 

The calculation of the $^{19}$F/$^{20}$Ne ratio is 
is somewhat more complicated than a folding of the cross section
and fluence: \\
$\bullet$ When Ne is excited by $\sim$ 20 MeV through inelastic
neutrino scattering, it breaks up in two ways
\begin{eqnarray}
^{20}\mathrm{Ne}(\nu,\nu')^{20}\mathrm{Ne}^* &\rightarrow& ^{19}\mathrm{Ne} + n 
\rightarrow {}^{19}\mathrm{F} + e^+ + \nu_e + n \nonumber \\ 
^{20}\mathrm{Ne}(\nu,\nu')^{20}\mathrm{Ne}^* &\rightarrow& ^{19}\mathrm{F}
+ p 
\end{eqnarray}
with the first reaction occurring half as frequently as the 
second.  The sum of these two channels is the
1/300 yield mentioned above. \\
$\bullet$ The subsequent nuclear processing determines whether the $^{19}$F
survives.  In the first 10$^{-8}$ seconds the coproduced neutrons in
the first reaction react via
\begin{equation}
^{15}\mathrm{O}(n,p)^{15}\mathrm{N}~~^{19}\mathrm{Ne}(n,\alpha)^{16}\mathrm{O}~~
^{20}\mathrm{Ne}(n,\gamma)^{21}\mathrm{Ne}~~^{19}\mathrm{Ne}(n,p)^{19}\mathrm{F} 
\end{equation}
with the result that about 70\% of the $^{19}$F produced via
spallation of neutrons is then immediate destroyed, primarily
by the $(n,\alpha)$ reaction above.  In the next $10^{-6}$ seconds
the coproduced protons are also processed
\begin{equation}
^{15}\mathrm{N}(p,\alpha)^{12}\mathrm{C}~~^{19}\mathrm{F}(p,\alpha)^{16}\mathrm{O}~~
^{23}\mathrm{Na}(p,\alpha)^{20}\mathrm{Ne} 
\end{equation}
with the latter two reactions competing as the primary proton
poisons.  This makes an important prediction: stars with high Na
abundances should make more F, as the $^{23}$Na acts as a proton
poison to preserve the produced F.\\
$\bullet$ A final destruction mechanism is the
heating associated with the passage of the shock wave.  Fluorine
produced prior to shock wave passage can
survive if it is in the outside half of the Ne shell.  The reaction
\begin{equation}
^{19}\mathrm{F}(\gamma,\alpha)^{15}\mathrm{N}
\end{equation}
destroys F for peak explosion temperatures exceeding $1.7 \cdot 10^9$K.
Such a temperature is produced at the inner edge of the Ne 
shell by the shock wave heating, but not at the outer edge.\\

If all of this physics in handled is a network code that
includes the shock wave heating and F production both before and
after shock wave passage, one finds~\cite{wh}
\[ \begin{array}{cc} \underline{[^{19}\mathrm{F}/^{20}\mathrm{Ne}]/
[^{19}\mathrm{F}/^{20}\mathrm{Ne}]_\odot} ~~&~~ \underline{T_{\mathrm{heavy}~\nu} \mathrm{(MeV)}} \\
0.14 ~~&~~ 4 \\ 0.6 ~~&~~ 6 \\ 1.2 ~~&~~ 8 \\ 1.1 ~~&~~ 10 \\ 1.1 ~~&~~ 12 \end{array} \]
for a progenitor star of solar metallicity.
One sees that the attribution of F to the neutrino process argues
that the heavy flavor $\nu$ temperature must be greater than 6 MeV,
a result theory favors.  One also sees that F cannot be overproduced
by this mechanism: although the instantaneous production of F
continues to grow rapidly with the neutrino temperature, too
much F results in its destruction through the $(p,\alpha)$
reaction, given a solar abundance of the competing proton poison
$^{23}$Na.  Indeed, this illustrates an odd quirk: although 
in most cases the neutrino process is a primary mechanism, one needs
$^{23}$Na present to produce significant F. Thus in this case the neutrino
process is a secondary mechanism. 

While there are other significant neutrino process products ($^7$Li,
$^{138}$La, $^{180}$Ta, $^{15}$N ...), the most important 
is $^{11}$B, produced by spallation off carbon.
A calculation by Timmes {\it et al.}~\cite{timmes} found that the combination of
the neutrino process, cosmic ray spallation and big-bang 
nucleosythesis together can explain the evolution of the light
elements.  The neutrino process, which produces a great deal 
of $^{11}$B but relatively little $^{10}$B, combines with the
cosmic ray spallation mechanism to yield the observed
isotope ratio.  Again, one prediction of this picture is that
early stars should be $^{11}$B rich, as the neutrino process 
is primary and operates early in our galaxy's history; the
cosmic ray production of $^{10}$B is more recent.  (We
return to this point below.)
There is hope that abundance studies will soon be able to descriminate
between $^{10}$B and $^{11}$B: as yet this has not been done. 

\section{The r-process}
Beyond the iron peak nuclear Coulomb barriers become so high
that charged particle reactions become ineffective, leaving
neutron capture as the mechanism responsible for producing
the heaviest nuclei.
If the neutron abundance is modest,
this capture occurs in such a way that each newly synthesized
nucleus has the opportunity to $\beta$ decay, if it is energetically
favorable to do so.  Thus weak equilibrium is maintained within
the nucleus, so that synthesis is along the path of stable 
nuclei.  This is called the s- or slow-process.  However a
plot of the s-process in the (N,Z) plane reveals that this
path misses many stable, neutron-rich nuclei that are known to
exist in nature.  This suggests that another mechanism is at
work, too.  Furthermore, the abundance peaks found 
near masses A $\sim$ 130 and A $\sim$ 190, which mark the closed
neutron shells where neutron capture rates and $\beta$ decay
rates are slower, each split into two subpeaks.  One set of subpeaks
corresponds to the closed-neutron-shell numbers N $\sim$ 82
and N $\sim$ 126, and is clearly associated with the s-process.
The other set is shifted to smaller N, $\sim$ 76 and $\sim$ 116,
respectively, and is suggestive of a much more explosive
environment where neutron capture is
rapid. 
  
This second process is the r- or rapid-process, characterized by: \\
$\bullet$ The neutron capture is fast compared to $\beta$ decay rates. \\
$\bullet$ The equilibrium maintained within a nucleus is established by $(n,\gamma) \leftrightarrow
(\gamma,n)$: neutron capture fills up the available bound levels in
the nucleus until this equilibrium sets in.  The new Fermi level
depends on the temperature and the relative $n/\gamma$ abundance.\\
$\bullet$ The nucleosynthesis rate is thus controlled by the $\beta$
decay rate: each $\beta^-$ capture converting n $\rightarrow$ p 
opens up a hole in the neutron Fermi sea, allowing another neutron
to be captured. \\
$\bullet$ The nucleosynthesis path is along exotic, neutron-rich
nuclei that would be highly unstable under normal laboratory conditions. \\
$\bullet$ As the nucleosynthesis rate is controlled by the $\beta$
decay, mass will build up at nuclei where the $\beta$ decay rates
are slow.  It follows, if the neutron flux is reasonably steady 
over time so that equilibrated mass flow is reached, that the
resulting abundances should be inversely proportional to these
$\beta$ decay rates.  Thus large abundances are expected at 
the shell closures, the ``waiting point'' nuclei where several
$\beta$ decays must occur before the shell gap inhibiting 
further neutron capture can be overcome.
  
The r-process requires exceptionally explosive conditions:
neutron densities in excess of $\sim 10^{20}$/cm$^3$, temperatures
of (1-3) $\times 10^9$K, and times on the order of one to a few
seconds.  Evaluating the $(n,\gamma) \leftrightarrow (\gamma,n)$
equilibrium for typical conditions yields neutron 
binding energies on the order of $\sim$ 30 kT, or about 2-3 MeV
below the neutron drip line.
After the r-process finishes (the neutron exposure ends)
the nuclei decay back to the valley of stability by $\beta$
decay.  This can involve some neutron spallation ($\beta$-delayed
neutrons) that shift the mass number A to a lower value.
But it certainly involves conversion of neutrons into protons,
which moves the r-process peaks at N $\sim$ 82 and 126
to lower N, clearly.  This shifted r-process peak combines with
the s-process peak to produce the double-hump distributions 
near neutron shell closures found in nature.
It is believed that the r-process can proceed to very heavy
nuclei (A $\sim$ 270) where it is finally ended by $\beta$-delayed
and n-induced fission, which feeds matter back into the
process at an A $\sim$ A$_{max}$/2.  Thus there may be important
cycling effects in the upper half of the r-process distribution.
  
What is the site(s) of the r-process?  This has been debated 
many years and still remains a controversial subject.
Both primary (requiring no preexisting metals) and secondary
(enriched in s-process elements) sites have been proposed.
Some of the suggested primary sites include
the neutronized atmosphere above the proto-neutron star in
a Type II supernova, neutron-rich jets produced in supernova
explosions or in neutron star mergers, and inhomogeneous big
bangs.  Secondary sites, where successful synthesis can result
for lower $\rho$(n), include the He and C zones in Type II
supernovae and the red giant He flash.

The balance of evidence favors a primary site, so one requiring
no preenrichment of heavy s-process metals.  
In particular, recent abundance studies~\cite{sneden} of
very metal-poor stars ([Fe/H] $\sim$ -1.7 to -3.12)
have yielded r-process distributions very much like that of our
sun (at least for Z $\gsim$ 56) (see Fig. 1).  In these stars
the iron content is variable.  This suggests that the ``time
resolution" inherent in these old stars is short compared to
galactic mixing times (otherwise Fe would be more constant).
The conclusion is that the r-process material in these stars
is most likely from one or a few local supernovae.  The fact
that the distributions match the solar r-process  
strongly suggests that there is some kind of
unique site for the r-process: the solar r-process distribution
did not come from averaging over many different kinds of
r-process events.  Clearly the fact that these old stars are
enriched in r-process metals also strongly argues for a 
primary process: the r-process works quite well in an
environment where there are few initial s-process metals.

\begin{figure}[htb]
\psfig{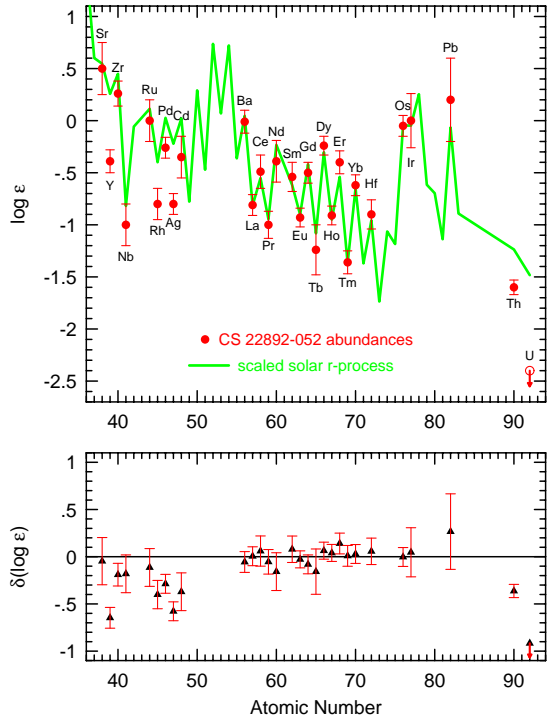}
\caption{Neutron-capture abundances in the ultra-metal-poor
([Fe/H] = -3.1) halo field giant star CS 22892-052 are plotted
as filled circles with error bars, along with a scaled solar
system r-process abundance curve (solid line).  In the bottom
panel, a differential comparison between individual elements
and the scaled solar system r-process abundance distribution
shows excellent agreement above Z = 56, but some deviations 
for lighter elements.  From Ref. 5.}
\end{figure}
  
It may be that these and similar data make certain
primary r-process sites, such as neutron star mergers, less probable.
The reasoning rests on the expected infrequency of neutron star
mergers (no more than 1/100th the rate of galactic supernovae), and
thus on the larger nucleosynthetic output required from such
r-process sites~\cite{qianw}.  Since the ejecta of neutron star mergers and
supernovae are expected to mix over similarly sized regions,
the former should produce a larger scatter of enrichments in
metal-poor stars.

These and other arguments have led many to suspect that 
core-collapse supernova may be the correct site.  There is
good theoretical support for this conclusion.
First, galactic chemical evolution studies indicate that
the growth of r-process elements in the galaxy is consistent 
with low-mass Type II supernovae in rate and distribution.
More convincing is the fact that modelers~\cite{woosley} have shown that the
conditions needed for an r-process (very high neutron densities,
temperatures of 1-3 billion degrees) might be realized in a
supernova.  The identified location is the last material blown off the 
supernova, the material just above the mass cut.  When
this material is initial at small $r$, it is a very
hot, neutron-rich, radiation-dominated gas containing neutrons
and protons, with neutrons dominating.  As it expands
off the star and cools, the material first goes through
a freezeout to $\alpha$ particles, a step that essentially
locks up all the protons in this way.
Then the $\alpha$s interact through reactions like 
\begin{eqnarray}
\alpha + \alpha +\alpha &\rightarrow& {}^{12}\mathrm{C} \nonumber \\
 \alpha + \alpha + n &\rightarrow& {}^9\mathrm{Be} 
\end{eqnarray}
to start forming heavier nuclei.  Unlike the big bang,
the density is sufficiently high to allow such three-body
interactions to bridge the mass gaps at A = 5,8.  The
$\alpha$ capture continues up
to A $\sim$ 80 in network calculations.  
The result is a small number of ``seed" nuclei,
a large number of $\alpha$s, and excess neutrons.  These 
neutrons preferentially capture on the heavy seeds to
produce an r-process.  Of course, what is necessary is to
have $\sim$ 100 excess neutrons per seed in order to 
successfully synthesize heavy mass nuclei.  While some calculations
come close to achieving this, the entropies tend to fall
short of what is needed.
An attractive aspect of this site is the amount of matter ejected,
about 10$^{-5} - 10^{-6}$ solar masses,
enough to produce the present galactic r-process metallicity
for a reasonable supernova rate.
  
It is clear that neutrino physics is an intimate part of the
r-process.  The supernova scenario described above is usually
attributed to material ejected by the protoneutron star's
neutrino wind.  This wind is also responsible for regulating
the essential proton/neutron chemistry of this material:
the reactions $\nu_e + n \leftrightarrow e^- + p$ and
$\bar{\nu}_e + p \leftrightarrow e^+ + n$ control this physics.
Nonstandard neutrino physics could be critical to the r-process.
An oscillation
of the type $\nu_e \rightarrow \nu_{\mathrm{sterile}}$ can alter 
the n/p ratio, as it turns off the $\nu_e$s that destroy neutrons
by charged-current reactions.  

The nuclear physics of the r-process tells us that the synthesis
occurs when the nucleon soup is in the temperature range of
(3-1) $\cdot 10^9$K, which, in the hot bubble r-process described above, corresponds to a freezeout radius of
(600-1000) km and a time $\sim$ 10 seconds after core collapse.
The neutrino fluence after freezeout (when the temperature
has dropped below 10$^9$K and the r-process stops) is then $\sim$
(0.045-0.015) $\cdot 10^{51}$ ergs/(100km)$^2$. 
Thus, after completion of the r-process, the newly synthesized
material experiences an intense flux of neutrinos.
This brings up the question of whether the neutrino flux could
have any effect on the r-process.  

\section{Neutrinos and the r-process}
Rather than describe the exotic effects of neutrino oscillations
on the r-process, mentioned briefly above, we will examine
standard-model effects that are nevertheless quite interesting.
The nuclear physics of this section -- neutrino-induced neutron
spallation reactions -- is also relevant to recently proposed
supernova neutrino observatories such as OMNIS and LAND.
In contrast to our first discussion of the $\nu$-process in
producing $^{19}$F and $^{11}$B, it is apparent that neutrino effects could be much
larger in the hot bubble r-process: the synthesis
occurs {\it much} closer to the star than our Ne radius of
20,000 km.  The r-process is completed
in about 10 seconds (when the temperature drops to about 
one billion degrees), but the neutrino flux is still significant
as the r-process freezes out.  The net result is that the
``post-processing" neutrino fluence - the fluence that can
alter the nuclear distribution after the r-process is completed -
is about 100 times larger than that responsible for fluorine
production in the Ne zone.  Recalling that 1/300 of the nuclei
in the Ne zone interacted with neutrinos, and noting that
the relevant neutrino-nucleus cross sections scale as A
(a consequence of the sum rules governing first-forbidden
neutrino cross sections), one
quickly sees that the probability of a r-process nucleus 
interacting with the neutrino flux is approximately unity.

Because the hydrodynamic conditions of the r-process are highly
uncertain, one way to attack this problem is to work backward
in time~\cite{hy}.  We know the final r-process distribution (what nature
gives us) and we can calculate neutrino-nucleus interactions
relatively well.  Thus from the observed r-process distribution
(including neutrino postprocessing) we can deduce
what the r-process distribution looked like at the
point of freezeout.  In Figs. 2 and 3, the ``real'' r-process
distribution - that produced at freezeout - is given by the 
dashed lines, while the solid lines show the effects of the
neutrino postprocessing for a particular choice of fluence. 

\begin{figure}[htb]
\psfig{bbllx=0.0cm,bblly=4.5cm,bburx=20cm,bbury=23.0cm,figure=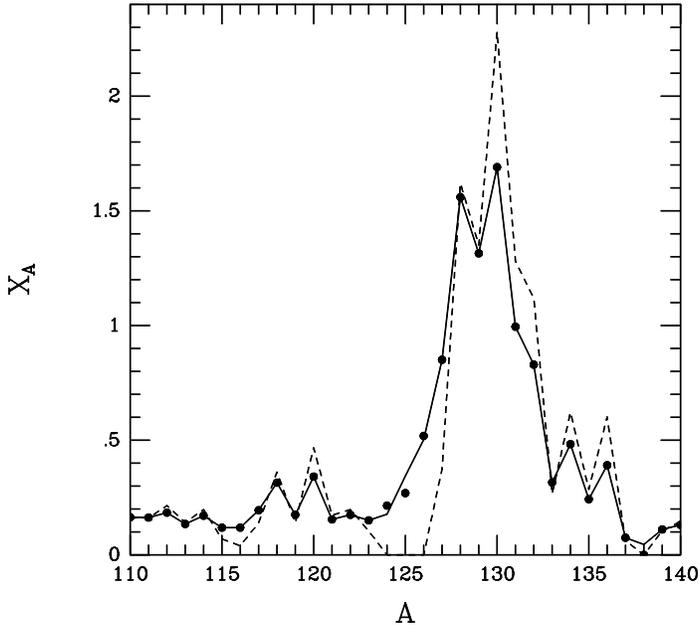,height=3.5in}
\caption{Comparison of the r-process distribution that would 
result from the freezeout abundances near the A $\sim$ 130 
mass peak (dashed line) to that where the effects of neutrino
postprocessing have been include (solid line).  The fluence 
has been fixed by assuming that the A = 124-126 abundances
are entirely due to the $\nu$-process.}
\end{figure}
  
One important aspect of the figures is that the mass shift is
significant.  This has to do with the fact that a 20 MeV 
excitation of a neutron-rich, weakly bound nucleus allows multiple neutrons
( $\sim$ 5) to be emitted.  
The relative contribution of the neutrino process is particularly
important in the ``valleys" beneath the mass peaks: the reason
is that the parents on the mass peak are abundant, and the
valley daughters rare.  In fact, it follows from this that the neutrino
process effects can be dominant for precisely seven
isotopes (Te, Re, etc.) lying in these valleys.  Furthermore
if an appropriate neutrino fluence is picked, these isotope
abundances are produced perfectly (given the abundance errors).
The fluences are
\begin{eqnarray}
   \mathrm{N} = 82~ \mathrm{peak:}~~~~~0.031 \cdot 10^{51} \mathrm{ergs/(100km)^2/flavor} \nonumber \\
   \mathrm{N} = 126~ \mathrm{peak:}~~~~0.015 \cdot 10^{51} \mathrm{ergs/(100km)^2/flavor} 
\end{eqnarray}
values in fine agreement with those that would be found
in a hot bubble r-process.  So this is circumstantial but 
significant evidence that the material near the mass cut of 
a Type II supernova is the site of the r-process: there is a
neutrino fingerprint. 

\begin{figure}[htb]
\psfig{bbllx=0.0cm,bblly=4.5cm,bburx=20cm,bbury=23.0cm,figure=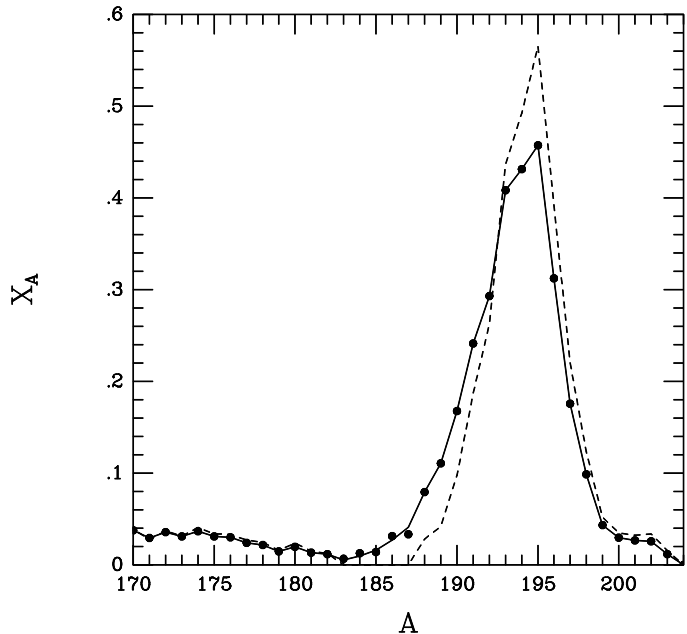,height=3.5in}
\caption{As in Fig. 5, but for the A $\sim$ 195 mass peak.
The A = 183-187 abundances are entirely attributed to the 
$\nu$-process.}
\end{figure}
  
\section{Neutrino Oscillations and Supernova Nucleosynthesis}
There are some intriguing connections between supernova
nucleosynthesis, the explosion mechanism, and neutrino
oscillations.  Several of these have to do with the
distinctive temperature hierarchy of supernova neutrinos 
mentioned earlier.  In contrast to solar neutrinos, where
detailed nuclear physics determines the neutrino spectrum,
the supernova neutrino temperature dependence on flavor is 
governed by very general arguments having to do with neutrino
couplings to matter, as we have noted.  While modelers differ
somewhat in their estimates of neutrino temperatures, there
is agreement that the heavy flavor neutrino mean energy is 
higher than that of the electron neutrinos, and that the $\nu_e$
temperature is lower than that of the $\bar{\nu}_e$s.  One
consequence is that neutrino oscillation signals in terrestrial
detectors could be quite obvious at the time of the next
galactic supernova.  For example, if $\nu_e$ events prove to
be substantially more energetic that $\bar{\nu}_e$ events,
the natural interpretation would be oscillations between heavy-flavor
and $\nu_e$ neutrinos, leading to an anomalously hot $\nu_e$
spectrum.

One important aspects of supernova neutrino oscillations is their
potential to probe the MSW mechanism over greatly extended 
parameter ranges.  The neutrinos have fixed spectra after they
decouple at the neutrinosphere, $\sim 10^{12}$ g/cm$^3$, a density
ten orders of magnitude greater than that at the core of the sun.
It follows that neutrinos with masses in excess of 100 eV
(thus $\delta m^2$ in excess of (100 eV)$^2$) will
experience an MSW crossing.  These crossings remain adiabatic --
that is, capable of converting neutrino flavor -- for mixing 
angles as small as $10^{-5}$, depending on the $\delta m^2$ value.
It follows that oscillations
unobservable by any other means could be revealed in supernovae.

Relevant to the present talk is the possibility that we will
not need to wait for the next supernova: oscillation effects 
might be deduced from their effects on nucleosynthesis.
For example, we have noted that a MSW oscillation between heavy
and electron flavors would lead to an unusually hot $\nu_e$
spectrum.  The proton/neutron chemistry of the hot nucleon soup
blown off the protoneutron star is governed by the competition
between the reactions
\begin{eqnarray}
\nu_e + n &\rightarrow& e^- + p \nonumber \\
\bar{\nu}_e + p &\rightarrow& e^+ + n.
\end{eqnarray}
As the oscillation leads to a hotter $\nu_e$ spectrum but does
not affect the $\bar{\nu}_e$s (which, for the usual mass hierarchy,
do not experience an MSW crossing), the first reaction is enhanced
while the second is unchanged.  The matter is thus driven proton-rich,
destroying any possibility of an r-process.  Thus, as Fuller 
has argued~\cite{fq}, a demonstration that the supernova ``hot bubble''
is the site of the r-process would impose very stringent constraints
on $\nu_e \leftrightarrow \nu_\mu/\nu_\tau$ oscillations.
The constraints address the entire range of cosmologically
interesting $\nu_\tau$ masses.

Another r-process connection arose from efforts to explain the
LSND, atmospheric, and solar neutrino results in four-neutrino 
schemes (three active and one approximately sterile).  One
such scheme involves a $\nu_\tau/\nu_\mu$ doublet at about 2 eV,
split in order to reproduce the atmospheric $\delta m^2$, and
a light $\nu_e/\nu_{sterile}$ doublet, split to reproduce the 
solar $\delta m^2$.  Such a scheme can have a salutory effect on
the r-process because of successive MSW crossings~\cite{caldwell}.  First the 
$\nu_\mu/\nu_\tau$ flux is removed by an oscillation 
with $\nu_{sterile}$; then a $\nu_e \rightarrow \nu_\mu/\nu_\tau$
oscillation can take place without a corresponding back
reaction.  With the $\nu_e$ flux reduced but the $\bar{\nu}_e$
unaffected, the matter can be driven neutron rich.  This occurs
at a radius where the increase in available neutrons helps the
r-process to succeed in producing the A $\sim$ 190 mass peak.

Oscillations could also influence our interpretation of the
abundances of the rare isotopes $^{10,11}$B, $^9$Be, and 
$^{6,7}$Li.  The neutrino process on $^{12}$C appears to
produce a great deal of $^{11}$B, consistent with its 
abundance.  It also produces significant $^7$Li, but very
little $^{10}$B, $^9$Be, and $^6$Li: neutral current neutrino
reactions generally do not impart sufficient energy to $^{12}$C
to populate the higher threshold channels corresponding
to these products.  The neutrino process produces a $^{10}$B/$^{11}$B
ratio of $\sim$ 0.05, while the true abundance ratio is
$\sim$ 0.25.  As a primary process, it predicts a linear growth
of boron with metallicity (e.g., Fe).

However, the textbook explanation for
the synthesis of these elements is the interaction of 
cosmic ray protons with $^{12}$C and $^{16}$O in the interstellar
medium~\cite{cr}.  As the reactions involve high-energy protons, $^{10}$B
and $^6$Li are readily produced: the $^{10}$B/$^{11}$B ratio
is $\sim$ 0.5, about twice that observed.  The cosmic ray mechanism
becomes more effective as the interstellar medium is enriched
in $^{12}$C and $^{16}$O and thus, as a secondary process, 
produces a quadratic growth in B with metallicity.  It is very possible
that nature uses a mixture of these two mechanisms.  If each
contributed about 50\% of the $^{11}$B, the correct $^{10}$B/$^{11}$B
ratio would result.

However recent studies of metal-poor stars show that the boron
grows linearly with Fe~\cite{duncan}: the production appears to be primary.
This has encouraged several efforts to reformulate the cosmic 
ray mechanism as a primary process, e.g., by accelerating 
$^{12}$C and $^{16}$O off a supernova on to target protons in
the interstellar medium.  (I believe the tasking of estimating
the production resulting from such a scenario is highly 
uncertain, however.)  Surprising quantities of $^9$Be have
also been observed in metal-poor halo stars.

Since the $\nu$ process is a primary process, one could consider
whether the calculated productions of $^{10}$B and other 
high-threshold products might have been underestimated.  It
is clear that the nuclear physics uncertainties affecting these
channels are much greater than in the case of $^{11}$B.
But another possibility that has not been explored is neutrino
oscillations.  In fact, if the neutrino masses have a standard
seesaw pattern, the atmospheric neutrino results suggest a
$\delta m_{13}^2$ somewhat below 0.01 eV$^2$, producing a
$\nu_e \leftrightarrow \nu_\tau$ MSW crossing near 10$^5$ g/cm$^3$.
This is outside the r-process region, thus leaving that
synthesis unaffected, but before the carbon zone at
$\rho \sim 8 \cdot 10^2 - 5 \cdot 10^3$ g/cm$^3$.  The hot
$\nu_e$ flux leads to enhanced production of $^{10}$B through
$^{12}$C($\nu_e,e^-)$, as well as increased $^7$Li through 
the burnup reaction $^{10}$B(p,$\alpha)^7$Be.  Numerical results
will be published soon~\cite{fh}.

\section{Summary}
The connections between neutrino nucleosynthesis and the supernova
mechanism are rather remarkable.  We have seen that neutrinos are
directly responsible for important synthesis. In turn this synthesis
can be exploited as a diagnostic of the explosion, e.g., as a monitor
of yet unmeasured heavy-flavor temperatures in the $\nu$ process,
and as a constraint on the explosion dynamics in the case of the
r-process.  (The neutrino fluence derived from the $\nu$ 
postprocessing ``fingerprint'' on the r-process constrains a 
product of the freezeout radius and the expansion rate.)

If one adds new neutrino physics to the equation, 
the occurence of a supernova ``hot bubble'' r-process places
important new constraints on the entire range of $\delta m^2$
relevant cosmologically.  
Four-neutrino scenarios postulated to account for the LSND,
atmospheric, and solar neutrino results can enhance r-process
production in the vicinity of the A $\sim$ 190 peak, and thus
could account for current underproductions in this region.
Finally, the relative mix of cosmic ray and $\nu$ process
synthesis of Li/Be/B would be affected by mixings of the $\nu_e$ 
governed by $\delta m^2$ near the atmospheric neutrino value.

The most interesting aspect of all of this is the impact new
abundances observations are having.  It gives one hope that
new $\nu$ physics might be learned from supernovae even before
the next flux of supernova $\nu$s hits the earth.

\section*{Acknowledgements}
I thank Scott Burles for providing, on behalf of the authors 
of Ref. 5, Fig. 1.
This work was supported in part by the US Department of Energy.

\end{document}